\documentstyle[aasms4]{article}

\begin{document}

\title{A far IR study of the CfA Seyfert sample: I. The data
\footnote{Based on observations
with ISO, an ESA project with instruments funded by ESA
Member States (especially the PI countries: France, Germany, the Netherlands
and the United Kingdom) and with the participation of ISAS and NASA.}}

\author{A. M. P\'erez Garc\'{\i}a\altaffilmark{1,2}, J. M. Rodr\'{\i}guez
Espinosa \altaffilmark{1}}
\affil{$^1$Instituto de Astrof\'\i{}sica de Canarias, E-38200
La Laguna (Tenerife), Spain }
\affil{$^2$
Isaac Newton Group, La Palma, Spain\nl
Electronic mail: apg@iac.es, jre@iac.es}


\begin{abstract}

We present mid and far IR ISO data of the CfA Seyfert galaxy sample. These data
allow a detailed study of the far IR Spectral Energy Distribution (SED) 
of these
galaxies. A Bayesian inversion method has been used to invert the SED of these
sources yielding two fundamental results, namely, that the mid and far IR
SED of Seyfert galaxies can be explained solely through thermal reradiation
of high energy photons by dust, and that this thermal emission is made up of
two or three different 
independent components, a warm, a cold and a very cold dust
component. These thermal components have been readily explained as produced
respectively
by warm dust heated by either the active nucleus or by circumnuclear
starburts, cold dust heated by star forming region in the galaxy disk, and
very cold dust heated by the general interstellar radiation
field. Comparisons between the parameters obtained from the analysis of the
IR SEDs (fluxes, temperatures, luminosities)  have been made. Our results
suggest that the emission in the mid IR is anisotropic and the differences
found between Seyfert 1 and Seyfert 2s can be explained with thin 
molecular tori models. 

\end{abstract}

\section{Introduction}

The origin of the IR emission from Seyfert galaxies has been a matter of 
discussion since the early work of  Rieke \& Low (1972);
Stein (1975); Rieke (1978); Neugebauer (1978),  and others. 
These authors found
an emission excess between 3 and 5~$\mu$m in Seyfert galaxies, that
originated a strong controversy between those 
who supported that the emission was
non thermal and those 
who defended that the excess was emission from dust heated by
the nucleus. The use of efficient instruments to observe at 10~$\mu$m, and
especially the pioneering work at the far IR carried out from the Kuiper
Airborn Observatory (KAO) showed the importance of the mid and far IR emission
to quantify the bolometric luminosity of Seyfert galaxies (Telesco \& Harper
1980; Smith et al. 1983). However, it was the IRAS satellite that provided
an extensive set of IR data for a large number of galaxies, 
from which it was shown
that Seyfert galaxies are indeed strong far IR emitters
(Rodr\'{\i}guez Espinosa, Rudy \& Jones 1987; Edelson, Malkan \& Rieke 1987; 
Spinoglio et al. 1995). 

The IRAS satellite was key to the understanding of
the importance of the IR emission to
the total luminosity of active galaxies. However the IRAS data alone are not
sufficient to clarify the nature of the IR emission, as there are 
measurements only at a limited number of wavebands preventing a good 
 definition of the shape of the Spectral Energy
Distribution (SED) at mid and far IR wavelength (see, e.g., Telesco
1988; 1990; Bregman 1990). 
A proper characterization of the mid and far SED 
is essential to understand the emission mechanisms that produce the
high output of Seyfert galaxies in the IR. Recent studies have
discussed the origin of the IR emission suggesting that it is of thermal 
origin. For example Giuricin, Mardirossian \& Mezzetti (1995) have
studied a complete sample of Seyfert galaxies at 10~$\mu$m and propose that 
the emission is due to thermal reemission by dust. Bonatto \&
Pastoriza (1997), based on color studies of IRAS data from diverse Seyfert
samples, find that the colors obtained can be explained with a combination of
dust heated by the nucleus plus cold dust in the host galaxy. Siebenmorgen et
al. (1997) show that the SED at IR and milimetric wavelengths can be modeled
assuming that the dust heated by a central source dominates the luminosity
 output of these objects. Maiolino et al. (1998) confirmed this last result 
finding, in their high resolution IR images of the Circinus galaxy, 
an unresolved source, with size $<$~1~pc, that is reprocessing the nuclear
 output via dust reradiation.
 Rigopoulou et al. (1997) have observed a sample of AGN 
in the CO mm line suggesting that the far IR emission of
Seyfert galaxies is thermal, based on three different
evidences: the correlation found between the far IR and the CO emission, the 
dependence of
the far IR emission to hydrogen molecular mass ratio with 
dust temperature, and the similarity of the profile shape of the 
CO and HI lines.

Another important issue is the understanding of the differences between the
two Seyfert types.
According to the unified models, Seyfert 2 nuclei are intrinsically similar
to Seyfert 1 nuclei, the differences observed being due
solely to geometrical effects. 
In Seyfert 2, neither the broad
line region nor the optical, UV and soft X ray continuum can be observed directly
because the central
region is obscured by intervening material in the line of sight. Some authors
 argue that this material forms a sort of disc or torus of molecular material. This disc
or torus is thought to be reponsible for the collimation of radiation and the
observed anisotropies, i.e., biconic structures in emission line images
(Simpson et al. 1997; Wilson et al. 1993) or highly collimated jets. It is
expected that at sufficiently long wavelengths the optical depth of the torus
would decrease and the differences between the two Seyfert types should
disappear. Several tests have been made to ascertain the presence of these
molecular tori (Heckman 1995; Pier \& Krolik 1993).
 These obscuring tori have been theoretically modeled by
Pier \& Krolik (1992) and Granato \&
Danese (1994) among others, predicting that the mid IR optical depth  is still
considerable, thus it should be expected that Seyfert 1s are more luminous than
Seyfert 2s in the mid IR.

In this work, we make an attempt to understanding the origin of the 
mid and far IR emission from Seyfert
galaxies through the study of their spectral energy distributions (SED).
We present ISO data of the entire CfA Seyfert sample (Huchra \& Burg
1992), consisting of 25 Seyfert 1 and 22 Seyfert 2, plus a LINER. 
 Section 2 describes the
observations. Section 3 presents the
separation of the Spectral Energy Distributions in thermal components
by means of the Inverse Planckian Transform.
 Section 4 describes the thermal emission components obtained from the
inversion and in Section 5 we perform a statistical analysis of the 
parameters obtained and discuss the differences between the two Seyfert types.

As the CfA
Seyfert sample is a complete sample of Seyfert galaxies we expect that
 the results
obtained here are statistically significant for all Seyfert galaxies and
certainly suggestive for other classes of AGNs.

\section{The data}

Observations of the 
CfA Seyfert sample have been carried out with the
Infrared Space Observatory (ISO; Kessler et al. 1996) 
through filters at 16, 25, 60, 90 120, 135, 
180 and 200$\mu$m. NGC~1068 was 
also observed at 4.9, 7.3 and 11.5~$\mu$m.
The filter set was chosen to achieve good coverage 
 in wavelength while producing a good sampling of the SEDs.
The ISOPHOT instrument (Lemke et al. 1996) was used in
 the PHT-P and PHT-C configurations, with the 
P1, P2, C100 and C200 detectors. P1 and P2 are single element
photodiodes.
 C100 is an array of 3x3 pixels, each one projecting onto 45
arcsec in the sky, while C200 is a 2x2 pixel array, each pixel projecting
 onto 89.4 arcsec in the sky.
 Integration times were calculated from the signal/noise ratio estimates
produced by the ISOPHOT simulator based on interpolations and extrapolations
of the IRAS data. The PHT-P observations were done in chopping mode through a
120 or 180 arcsec aperture depending on the size of the objects. 
The PHT-C
observations were done in staring mode. In this case, to set the background
level, an
empty area adjacent to the source was observed prior to the actual
source and with identical instrumental settings.
For NGC~1068 we have done maps with C100, at 60 and 105~$\mu$m, moving by half
a pixel (23'') between two contiguous map positions.
Four additional 
objects (NGC~3079, NGC~3227, NGC~4051 and NGC~4151) have been  mapped
at 90~$\mu$m with the C100 array. Details of the mapping observations 
of these four galaxies are given in
P\'erez Garc\'{\i}a, Rodr\'{\i}guez Espinosa \& Fuensalida (2000). 

The data reduction was done with the help of the PHT-Interactive Analysis
(PIA) V.7.0 tool. PIA first deglitchs the data eliminating the 
cosmic rays 
and others spurious effects, then corrects for  non-linear effects
taking into account both the dynamic range of the detectors and the flux
level of the observed sources. To determine the signal PIA linearizes 
the integration 
ramps, and corrects for detector drifts. The background substraction
is achieved by repeating the PIA reduction process for the background
measurements files, which are then substracted from the object files. 
It is also necessary to correct for the PSF fraction that is not seen by
the detector (see Lemke et al. 1996). 
This factor varies with wavelength, hence the actual correction factor
 for each filter  is built-in in  a PIA table.  
The flux calibration is performed using the standard calibration of PIA
7.0. 
For the PHT-P measurements the process is similar except
that the background substraction is performed automatically from the chopped
measurements and the flux calibration is done through the use of the internal
FCS (Fine Calibration Source; Lemke et al. 1996) 
since it provides a better match with 
the actual intensity of the
sources. Note that the P detector response is very dependent on the 
intensity of the source being observed.

The photometric uncertainties are small varying between  1\% for
the objects with higher signal/noise ratio and  15\% for the weakest
objects. The final uncertainties are however dominated by the
uncertainties in the flux calibration. 
The calibration factors have kept improving as the detector and instrument
characteristics have become better known, although there are still 
some residual effects (non linearity of detectors) that are yet not adequately
modelled. As of this writting, the 
uncertainty of the calibration is equal or better than 30\%.  
Therefore, throughout this paper,  we adopt a 
conservative uncertainty of 30\%. 

The fluxes for the observed objects are given in Table \ref{tab1} and 
\ref{tab2}. 
Fluxes are in Janskys. Some objects could not be observed 
in some or all of the filters for various reasons. 
These are also given in Table 1.

Figure \ref{fig1}  shows the Spectral Energy Distributions (SEDs) 
of the observed objects. The ISO data have been plotted together with   
 IRAS data (Edelson, Malkan \& Rieke 1987) at 12, 25, 60 and 100$\mu$m
showing that the agreement  
 of the ISO and IRAS data is very good in most cases.


\placetable{tab1}
\placetable{tab2}

\section{The inversion of the SEDs}

It can be seen in fig. 1 that the SEDs  describe
a well defined energy range with a steady increase from the near IR on to a 
maximum between 90 and 135$\mu$m and 
the start of a decline toward longer wavelengths.
Several results like those mentioned in the introduction
 point to the mid and far-IR emission of Seyfert galaxies being of
thermal origin. Furthermore, we have shown that the SED of a few Seyfert
galaxies can be explained as emission produced by two emissivity 
weighted blackbodies (Rodr\'{\i}guez Espinosa et al. 1996). 

\placefigure{fig1}

However, rather than proceeding with a plain fit of a ``ad hoc''number of
black bodies, 
we have preferred to use an inversion method to analize the SEDs of the galaxies in the
sample. This has the advantage that no assumptions have to be made as to the
number or location of the sources responsible for the observed spectrum. In
particular, we have used an Inverse Planckian Transform algorithm, that
employs an emissivity ($\epsilon \propto \lambda^{-1.5}$) weighted Planck function kernel to
switch from frequency space to the
temperature domain, hence revealing the temperature distribution of the 
sources that originate the observed
SEDs. The method applies  Bayes theorem of conditional probability and the
Richardson-Lucy iteration algorithm which converges quickly 
to a optimum result.
To increase the number of data points used in the inversion algorithm, 
whenever available, we have
added the four IRAS band fluxes at 12, 25, 60 and 100~$\mu$m as given by
Edelson, Malkan \& Rieke (1987). Furthermore, to 
avoid boundary convergence problems
in the inversion algorithm we have used 10~$\mu$m ground-based data from
several authors (Rieke 1978; Edelson, Malkan \& Rieke 1987; 
Maiolino et al 1995), and 1.3~mm upper limit data from
Edelson, Malkan \& Rieke (1987). These additional data have been used
solely for the purpose of constraining the inversion algorithm 
at the borders of the wavelength range of interest. Details of 
this method are given in Salas (1992) and in P\'erez
Garc\'{\i}a, Rodr\'{\i}guez Espinosa,  \& Santolaya Rey (1998) and are not
repeated here.

Figure \ref{fig2} shows the results obtained after application of the
Inverse Planckian Transform to the SEDs of the CfA Seyfert galaxies observed. 
  For each object, the upper pannel in Figure \ref{fig2} plots the ISO
data (triangles) and the IRAS data (filled squares) and the best fit to the
mid and far IR data in heavy black, with the different spectral components
contributing to the fit printed in dashed lines; the bottom pannel shows
the temperature spectrum that produces these components. 

\placefigure{fig2}

It is important to note 
that the temperature spectra obtained in this way are continuous. 
 Further, the spectral temperatures of the dust grains group themselves
in discrete and well defined features
that we have called components. These components can be described by their
peak 
temperature. Figure \ref{fig2} shows that 34 out of the 40 galaxies for which
we have obtained good temperature spectra
show three temperature components: a warm
component with central temperature
 T$\sim$150~K, a cold component with T$\sim$40-70~K, and 
a colder component with T$\sim$15-25~K. The remaining 6 objects show only two
components, although in NGC~1068 when the short wavelength data are added a
third component is recovered.
Table \ref{tab3} shows the peak temperatures of these 
thermal components for each
object. This table also gives the spectral and morphological type for each
object.
 These results 
confirm the analysis made before for a few objects of the
same sample (P\'erez Garc\'{\i}a, Rodr\'{\i}guez Espinosa,  \& Santolaya Rey
1998).

The fits obtained from the inversion are, in general, very good. In some
cases, weak objects or objects with low signal to noise 
ratios, the uncertainty in the inversion is
higher. For some objects, we have rejected some measurements (the flux at
120$\mu$m of NGC~4235, and the fluxes at 90$\mu$m of Mrk~270, Mrk~461  and
1614+35) because these data points do
 not follow the shape of the SEDs defined by the
rest of the measurements. In all of 
these cases the rejected measurement
 is very noisy.
Four other objects (Mrk~590, Mrk~573 and Mrk~841) have not been 
measured at long wavelengths ($>$90$\mu$m), because they are 
below the sensitivity limit for detection with ISO.

Nevertheless, the method still finds three components for three out of the
except for NGC~1068, that will be analized in a separate
section. four objects. For Mrk~590, the inversion does not converge.

In four of the objects, the warm component peak temperature obtained from the
inversion is not
high enough to fit the data. These four objects are Mrk~335, NGC~5273,
Mrk~1243 and IZw1. Both in 
Mrk~335 and in NGC~5273 it can be due to the low signal to noise
ratio of the data. 
 NGC~5273 shows a discrepancy between the IRAS value at 12~$\mu$m
and the ISO value at 16~$\mu$m (again very likely due to the low  
signal to noise ratio of the ISO data point) that prevents a good fit to the data.
Mrk~1243 has only three data points in the hot part of the spectrum. 
IZw1 is a special case, since the short wavelength data are  
of good signal to noise ratio, however it is
not detected at 200~$\mu$m and the flux at 180~$\mu$m is rather low. 
The fall of the 
signal at longer wavelengths is fast enough to explain that the object is
not detected  at 200~$\mu$m. Figure \ref{fig2} shows that the SED of IZw1
if made of just two components does not account 
for all of the observed emission.
If we substract this initial two-component fit 
from  the actual SED, the residual seems to correspond
to yet one third hotter thermal component. 
The inversion method does not succeed at 
identifying completely this third component, due to the temperature of this
third component being too
 high, therefore only a small fraction of the energy within this third
component falls within
 the range of wavelength studied.
The shape of the residual is however 
similar to the other two components,
a clear hint of the existance of yet another thermal component with a peak
temperature of $\sim$300~K. This temperature is the highest temperature found
among  all of the SEDs.
The cold and very cold components also have peak temperatures higher than the
 corresponding 
 components of the SEDs of the rest of the CfA sample. IZw1 is one of the
more distant and brighter Seyfert galaxies known. Its different
spectroscopic characteristics of line width and line ratios 
were pointed out by Osterbrock and Pogge (1985) and Goodrich (1989). Recently,
several authors have pointed out the different nature of IZw1.
 For example, the optical and X-ray spectra are steeper in
IZw1 than in normal 
 Seyfert 1 and 2. Indeed IZw1 shows fast and large
amplitude variability in its X-ray emission (see Halpern \& Moran 1998 and
references there in), and is possibly better associated with the 
broad line quasars
(Boroson \& Meyers 1992; Lawrence et al. 1997). 

\section{The dust components}

 
The inversion method assigns to each dust
component a range of temperature, i.e., the dust grains have temperatures
ranging between a minimum and a maximum temperature.
 This is indeed an expected and physically meaningful result, as the
dust closer to the source or sources of radiation will be warmer than the dust
farther away. The existence of temperature components indicates that there are well defined sources with well
defined temperature profiles which must be the result of well defined
physical scenarios existing in these objects. 
From the temperature spectra, we can calculate the flux that each dust component
contributes to the total mid and far IR emission. For each component, the
flux is obtained by integration of the temperature spectrum over the
relevant range of temperatures:

\begin{equation}
F_i = \int_{Tmin_i}^{Tmax_i} \Psi(T) dT \;\;\; i=1,2,3
\end{equation} 

where T$_{min_i}$ and  T$_{max_i}$ are the extreme values of the temperature
range that defines each component, and $\Psi(T)$ is the temperature 
distribution obtained from the inversion process.
 Table \ref{tab4} shows the fluxes obtained in this way  
for each component and the total flux for each object. 
 The comparison between the far IR IRAS fluxes (FIR), as
calculated from the values at 60 and 100~$\mu$m (Lonsdale et al. 1984) and
the sum of the cold and very cold components calculated here is in good
agreement.  
The luminosities for each of the components are shown in Table \ref{tab5}. We have used
$H_0$ = 75 Km s$^{-1}$ Mpc$^{-1}$ throughout this paper. 

The key point now is wheather these emission components can be physically
 explained within a sensible scenario. In what follows we
 review the pieces of evidence that we have for explaining each of components:

 a)  In a previous paper Rodr\'{\i}guez Espinosa \& P\'erez Garc\'{\i}a
(1997) were able to use optical R band images of a subset of low redshift
Seyferts of the present sample to separate the fluxes from the
central region from those of the galaxy disks. We found a very good
correlation between the ratio of the extended to the compact R band fluxes
and the ratio of the cold plus very cold component to the warm emission
component. The conclusion was that the  
  warm emission component is related with the central regions of the galaxies while
the cold and the very cold emission had to originate in the disk of these
galaxies.

 b)  Furthermore, a correlation has been found
 between the flux produced by the warm component and the flux in the high ionization coronal
 lines fluxes like [OIV]$\lambda$25.9 and [NeV]$\lambda$14.3$\mu$m
 (Prieto, P\'erez Garc\'{\i}a \& Rodr\'{\i}guez Espinosa 2000), indicating
that the warm component must be heated by the nucleus of these galaxies.

 c) In another recent paper, P\'erez Garc\'{\i}a et al. (2000) have shown
that based on 90$\mu$m ISO maps of four nearby Seyferts, the 90$\mu$m
emission is physically extended up to radii similar or larger than those seen
in optical images of these same galaxies. Furthermore, the extension of this
90$\mu$m  emission has been characterized and its extension, scale length and
surface brightness profiles are typical of normal galaxy disks. 

Based on the above, a scenario arises in which the warm component is
associated with dust heated by  radiation coming
from the nuclear or circumnuclear regions of these galaxies, while the cold
and very cold dust must be heated by process occurring in the galaxy disk. Danese et
al. (1992) also conclude
that the mid IR emission (10-25$\mu$m) is dominated by the nucleus or the
circumnuclear region. Further support to this scenario comes from the following:

\begin{itemize}
\item Warm dust. Its characteristic peak temperature 
 is in the range 120-170~K, a range of temperatures warmer tham is normal of
dust in typical starforming regions. The nuclear origin of the radiation
responsible for the heating of the warm dust was already indicated by Rudy
(1984) who found a correlation
between the [OIII]$\lambda$5007 emission line flux and the 10~$\mu$m
emission for a sample of quasars, Seyfert galaxies and radiogalaxies. This
 author suggested that the dust responsible for the 10~$\mu$m 
emission is mixed
 with the ionized gas of the Narrow Line Region (NLR) that produces the
 [OIII]$\lambda$5007 emission line. More recently, 
 Giuricin, Mardiossian and Mezzetti (1995) support
  the idea of the nuclear heating of this warm dust, based on
 10~$\mu$m small aperture observations of a sample of 100 galaxies. They
 found that the 10~$\mu$m emission correlates very well with the 25~$\mu$m IRAS
 emission, while the correlation is poor with the 60~$\mu$m emission. 
 From a different perspective Heckmann et al. (1997) have found in 
the Seyfert 2 galaxy
 Mrk~477 a strong starburst with very warm  dust in an scale of a few hundred parsecs,
 that confirms the idea that the warm dust can also be heated by nuclear and
 circumnuclear starformation regions.

\item Cold dust. Peak temperatures for this component range between 40 and
70~K, a range of temperatures that is typical of dust in regions of
starformation.
 Note that cold dust is present in mostly all classes of galaxies, 
including normal
and starburst galaxies (Knapp et al. 1996; Chini, Kruger \& Kreysa 1992;
Klaas et al 1997; Walterbos \& Schwering 1987). In all these galaxy types, 
the heating of the dust is
produced by OB stars in star formation regions in the galaxy disks. The higher
temperatures corresponding to higher recent starformation rates (Young et al.
1989)

\item Very cold dust. Peak temperatures for this component range
 from around 15 to 25~K, temperatures that are normal of dust heated
by the general interstellar radiation field. This very cold dust
is tipically composed of
 big grains in thermal equillibrium with the interstellar
radiation field and has been observed in normal spiral galaxies
(see, e.g., Walterbos \& Greenewalt 1996; Walterbos \&
Schwering 1987; Cox, Kruger \& Metzer 1986). For example, Cox, Kruger \&
Mezger (1986) predict that the dust of the Galaxy is at a temperature of 
around 15-20~K, and
Walterbos \& Schwering show that in the disk of M31 there is very cold dust
at 21~K. 
\end{itemize}

\subsection{NGC 1068\label{n1068}}

NGC~1068 has been observed at a wider spectral range than the rest of the
sample, 11 filters between
4 and 200~$\mu$m have been employed.
NGC~1068 is indeed one of the best studied Seyfert 2 galaxies in all spectral
ranges. Morphologically it is clasified as SAb, and was the first
galaxy in which broad emission lines in polarized light were found 
(Miller \& Antonucci
1983; Antonucci and Miller 1985). In the IR range, Tresh-Frenberg et
al. (1987) have found that the emission between 5 and 20~$\mu$m presents an
elongated structure in the nucleus, that they explain as thermal emission
by dust heated by both 
the active nucleus and an active starforming region near
(1''-2'') the nucleus. The comparison between 10$\mu$m images and HST
images (Cameron et al. 1993) indicates a correlation between the hot dust
emission and the Narrow Line Region.  These autors developed a model that
replaces the molecular torus by giant molecular clouds of gas and dust
that attenuate the broad lines.  Braatz et al. (1993) found a correlation
between a 12.4~$\mu$m image and both optical continuum and [OIII] images; they
 also observed that the IR emission is aligned perpendicular to the 
torus plane
and infered that the extended IR emission is thermal radiation by dust in
molecular clouds heated by collimated nuclear emission, without ruling
out a possible contribution by dust in the molecular torus. 
Telesco \& Harper (1980) have studied the far IR emission of NGC~1068 with
KAO (Kuipern Airborn Observatory) observations between 30 and 300~$\mu$m and
have found two thermal components with temperatures of $\sim$36~K and
$\sim$115~K.   

We have applied the inversion method explained before 
in the same spectral range that
we have used for the rest of the objects in the sample, i.e., between 12 and
200~$\mu$m. Figure \ref{fig3} shows the results obtained, i.e., 
 two thermal components, a cold one with 
peak temperature of T$\sim$36~K and a warm one with T$\sim$115~K in the
peak. It remains however to extend the fit to the shorter wavelength data
which we have for NGC~1068 alone. When 
 we apply the inversion method to the full range observed (4-200~$\mu$m)
 the SED of NGC~1068 appears now as 
the sum of three thermal components
(see Figure \ref{fig3}). Two of them can be identified with the two thermal
components previously 
found before without considering the 4-12~$\mu$m range. The peak
temperatures of the three components are now 
 29, 110 and 278~K. The differences between
the flux contained in each of the two components common to both inversions 
is less than 7\%. 

\placefigure{fig3}

The three components found in NGC~1068 are explained as in 
the rest of the objects of the CfA
sample. The cold component is produced by dust in the host galaxy, the
warm component corresponds to dust heated by the active nucleus and/or
circumnuclear starforming regions. The dust at higher temperature 
 is also heated by the
active nucleus, and should either be located
 in the inner part of the torus or be heated by active
starforming regions placed in a radius $<$100~pc from the active nucleus
(Gonz\'alez Delgado et al. 1998). 
The flux enclosed under 
 the component with T$\sim$278~K is F = 9.17 10$^{-9}$ erg cm$^{-2}$
s$^{-1}$, and the corresponding luminosity L = 2.40 10$^{45}$ erg s$^{-1}$, while the total
IR luminosity of NGC~1068 is 8.60 10$^{45}$ erg s$^{-1}$. Hence this
component represents 28\% of the total IR luminosity, which agrees
with the fraction of the total UV emission that is produced by circumnuclear
starforming regions in a sample of Seyfert 2 galaxies (Gonz\'alez Delgado et
al. 1998).  

\section{Energy balance: The cold emission component}

One possible test that 
can confirm the suggestion made before on the relation between
the cold emission component and the presence of
 starforming regions in the galaxy discs 
consists of testing the ratio of H$\alpha$ emission to far IR 
output. 
For instance in NGC~3079, the observed H$\alpha$ luminosity is
7.93 10$^{40}$ erg sec$^{-1}$ (Armus, Heckman \& Miley 1990), 
a figure that includes both the nuclear and extended H$\alpha$ luminosity.
 Moreover, with a ratio of extended to nuclear H$\alpha$ emission of
0.77 (Armus, Heckman \& Miley 1990), the extended H$\alpha$ emission amounts to 3.49 10$^{40}$ erg sec$^{-1}$,
 which shows the importance of the H$\alpha$ flux 
produced in star forming regions outside the central region in this object.
If we take as an average extinction the value measured 
by Hawarden et al. (1995) A$_V$=7.5 mag, the
extinction at $H_\alpha$ is A$_{H_\alpha}$=6.1 mag, and
the extinction corrected extended H$\alpha$  luminosity is 0.93 10$^{43}$ erg
sec$^{-1}$. This is to be compared with the IR output due to dust in
starforming regions that competes with the gas for high energy
 photons from massive
stars. In NGC~3079 this is what we have called the cold emission
(P\'erez Garc\'{\i}a, Rodr\'{\i}guez Espinosa \& Santolaya Rey 1998), 
which amounts to 3.52 10$^{44}$ erg
sec$^{-1}$, or a ratio of H$\alpha$ to cold FIR emission of 0.03. 
This value is in agreement with
the expected value in regions where massive stars are forming (Devereux \&
Young 1990). Hence for NGC~3079 there is good agreement between the
H$\alpha$ output from  disk HII regions and the
cold IR emission, which confirms the results in P\'erez
Garc\'{\i}a et al. (1998) and in this work indicating that 
the cold component emission in Seyfert galaxies originates in
starforming regions wjthin their disk. 

In the case of NGC~4051 we consider the ratio between H$\alpha$
and far IR luminosity  in two limiting cases. First, we consider
the total integrated H$\alpha$ luminosity, 4.34 10$^{41}$ erg sec$^{-1}$
(Romanishin 1990), which after correction for extinction (A$_V$= 0.24,
Ho et al. 1997) becomes 5.22 10$^{41}$ erg sec$^{-1}$. Hence, the ratio of
total H$\alpha$ to cold far IR emission (0.25 10$^{44}$ erg sec$^{-1}$) 
 is 0.02. This is an upper limit as
the total H$\alpha$ emission includes the nuclear emission which to a certain
extent can be due to non-thermal processes associated with the AGN. On the
other hand, if we consider the H$\alpha$ emission coming exclusively from the
star-forming regions in the disk of NGC~4051, which according to Gonz\'alez
Delgado et al. (1997) is 26\% of the total emission we get a ratio of
extended H$\alpha$ to cold far IR emission of 0.006. This is a
lower limit as some of the H$\alpha$ flux that we have assumed nuclear is
very likely due to extended circumnuclear emission that will also contribute
to the cold far IR emission. The actual value for the ratio of 
extended H$\alpha$ flux
to cold far IR emission ranges therefore between 0.006 and 0.02, a value that
is well within the range given by Devereux \&
Young (1990) as typical of normal HII regions in the disk of spiral galaxies.

In the case of NGC~5033 the integrated H$\alpha$ flux is 3.98 
10$^{41}$ erg sec$^{-1}$ (Kennicutt 1983), which after correcting for
extinction (A$_V$= 1.48, Ho et al. 1997) becomes 1.20 10$^{42}$ erg
sec$^{-1}$. The H$\alpha$ emission to cold far IR
emission ratio  is 0.02, which is again to be taken as
an upper limit  
given that the total H$\alpha$ flux includes the nuclear emission
which we can not separate because we do not have the necessary data, 
However this upper limit
is similar to that obtained for NGC~4051. 

As for the rest of the sample a similar exercise with H$\alpha$ fluxes
obtained from the literature produce similar results. In very few instances  have
we been able to separate the extended from the nuclear H$\alpha$ emission. For
these cases the ratio of extended H$\alpha$ emission to cold far IR emission
ranges between 1. 10$^{-5}$ for NGC~5929 and 0.02 for Mrk~270. In most cases
this separation between extended and nuclear emission 
has not been possible and hence ratios
between the total H$\alpha$ emission and the cold far IR output have been
obtained. These range between 1.5 10$^{-4}$ (Mrk~766) and 2. 10$^{-3}$
 (NGC~3516). In all cases it can be seen that the ratios are within those
expected if
massive stars are responsible for the ionizing photons as well as for the
heating of the dust. We therefore conclude that the cold emission component
that appeared naturally from the inversion process is related
to radiation from dust in starforming regions in the galaxy discs.

  
\section{Seyfert type differences}
 
The current understanding of Seyfert galaxies assumes that the differences
between the two Seyfert types are due to geometric factors in the nuclear
region. If this is so we should look for differences in the warm emission
component between the Seyfert types.
Let us therefore turn our attention to the warm dust component that we have suggested  it is
directly related to the nuclear and circumnuclear dust emission. If there is
a 
molecular torus obscuring the nuclear region
and this is sufficiently thick (as the models predict, see e.g. Pier
\& Krolik 1992; Granato \& Danese 1994) we shall observe
differences in the warm component emission produced by each
 of the two Seyfert types. 

In order to compare the parameters obtained from the inversion of the
SEDs for the different Seyfert types, 
we establish two different groups of objects within the CfA sample:
Seyfert 1s, including galaxies classified as Seyfert 1.5, and Seyfert 2s, 
including Seyfert 1.8 and 1.9. Therefore, the sample gets divided into
22 Seyfert 1 and  24 Seyfert 2. If we drop those objects not observed or
those whose data do not allow obtaining a well defined warm component, 
the sample with adequate data for
this analysis consists of 18 Seyfert 1 galaxies and 22 Seyfert 2 galaxies.

If the dust responsible for the warm emission component is located within 
 the molecular torus that hides both the active
nucleus and the broad line region, we expect to find differences between some
 of the
parameters that define the warm component of the Seyfert 1 and 
2 objects. If however the dust is located outside the molecular torus, 
then the geometry
 does not play anymore a key role and we should not see 
differences between the two
 types of Seyfert galaxies.

\subsection{Temperature differences}

Regarding the temperature
distribution,  it should be realized that in both Seyfert classes we are
sampling dust emission produced under similar physical
conditions, probed with the same wavelength range, and 
heated in either case by the active nucleus or by circumnuclear starforming
regions. Therefore as temperature is an intensive quantity as opposed to 
extensive, we expect not to see any difference between the average
temperature of the warm component in the two Seyfert types. 

Figure \ref{fig4} shows the distribution of warm component temperature
for the two groups of Seyfert galaxies. It is seen that the 
median temperature of the Seyfert 1s 
is slightly higher than the median of the Seyfert 2s though the
differences are small, IZw1 excepted. As we pointed out 
 before, IZw1 is a special
object, its warm temperature being way far out of every other object 
in the sample. We have therefore dropped IZw1 from the statistics.
To understand the significance of the 
differences between the two distributons, we have applied a
Kolmogorov-Smirnov (KS) test. 
The characteristics of the distributions are:

\[ \begin{array}{lll}
\langle T \rangle _1&=  148 &\sigma = 12.\\
\langle T \rangle _2 &= 144 &\sigma = 14.\\
\end{array} \] 

The result of this KS test is that there is a 
78\% probability that the two
distributions are the same and so the differences between
them are not significant and both groups 
(Seyfert 1 and Seyfert 2) enjoy, as expected, the same 
temperature distribution.

\subsection{Fluxes and luminosities}

Recent studies of the mid IR emission from Seyfert galaxies claim
 that  Seyfert 2 galaxies are weaker than Seyfert 1s
(Heckman 1995; Maiolino et al. 1995; Giuricin, Mardirossian \& Mezzeti 1995;
Mulchaey et al. 1994). This result is interpreted within the framework of the
unified models as an anisotropy, resulting from the presence of a molecular
torus with a given optical thickness in the IR. To test this claim 
we have compared the fluxes and luminosities of the warm 
component of the Seyfert
SEDs to see if we find significant differences between Seyfert 1 and
Seyfert 2s.

The median values for the warm luminosities are:
 
\[ \begin{array}{lll}
\langle log L_{warm} \rangle_1  &= 44.7& \sigma = 0.3\\
\langle log L_{warm} \rangle_2  & =44.9& \sigma = 0.8\\
\end{array} \]

Statistically, the KS test shows that
there is a probability 33\% that both distributions are the same. 
Hence there are no significant differences between the Seyfert 1 and 2's
regarding the warm emission.

Turning now to the total far IR luminosities, i.e., the sum of the three
emission components, the distributiones are:

\[ \begin{array}{lll}
\langle log L_{IR} \rangle_1  &= 45.1& \sigma = 0.8\\
\langle log L_{IR} \rangle_2  & =44.9& \sigma = 0.7\\
\end{array} \]

In this case the distributions are the same with a significance level of
99\% therefore it can not be concluded that the Seyfert 1s are brighter than the
Seyfert 2s in the mid and far IR. 

We now turn to search wheather there are differences between
Seyfert 1 and Seyfert 2s in the
ratio of warm component to total luminosity. 
First, we refer to the ratios between the
warm and total fluxes. The mean and standard deviations of
these distributions are: 

\[ \begin{array}{lll}
\langle F_{warm}/F_{IR} \rangle_1  &= 0.42& \sigma = 0.17\\
\langle F_{warm}/F_{IR} \rangle_2  & =0.28 & \sigma = 0.15\\
\end{array} \]

Figure \ref{fig4} shows the flux ratio distributions. A KS test indicates
 that the distributions are different at the
99\% level of significance. This result  suggests
that the Seyfert 1s emit fractionally more than the Seyfert 2s in the mid IR
 (warm component).
This is consistent, for example, with the models of circumnuclear tori of
Pier \& Krolik (1992;1993) and of 
Granato \& Danese (1994), that predict anisotropy by absorption of nuclear
emission in the mid IR. 
However this result should be taken with care, 
as it could indicate either that the nuclear emission
is larger in the Seyfert 1s relative to their total FIR flux 
or that the contribution of the 
host galaxy is stronger in the Seyfert
2s. To discriminate between these two possible explanations we must 
normalize the
infrared fluxes with an isotropic property of the galaxies in the sample,
i.e., with fluxes emitted at long enough wavelengths that they are not
suspect of suffering extinction and thus do not depend on the
geometry of the sources.
Other authors have used [OIII]$\lambda$5000~$\AA$ fluxes, hard X-ray 
fluxes or radio fluxes
(see, for example, Mulchaey et al. 1994). However, the 
[OIII] $\lambda$5000~$\AA$ fluxes can be
affected by absorption due to dust in the NLR. We prefer to
 use  20~cm radio
emission fluxes to normalize the IR flux, 
since the radio emission is not affected
by selective extinction. We have used integrated radio data from 
Edelson (1987). 
These data consists of VLA
observations at 1.46~GHz (20~cm) with a bandwidth of 45~MHz. The FWHM
beamwidth used is 1.5 arcmin, directly
comparable with our ISO data. 

The distributions of the radio normalized warm IR fluxes show the following
characteristics:

\[\begin{array}{lll} 
   \langle  log (F_{warm}/F_{20cm})\rangle_1 &= 6.5 & \sigma= 0.3\\
   \langle  log (F_{warm}/F_{20cm}) \rangle_2 &= 6.1 & \sigma= 0.3 

    \end{array}\]\\
 
KS tests indicate that 
both distributions are different at a significance level of 99.9\%. Therefore,
the warm flux is indeed higher in Seyfert 1s than it is in Seyfert 2s, and it can be
 concluded that the warm emission from Seyfert 2s is affected by dust
extinction to a larger extent that in the Seyfert 1 galaxies. 


This result
suggests that at shorter wavelengths (mid IR)
the emission is still anisotropic, 
in agreement with the molecular torus models of Pier \& Krolik (1992;1993)
and Granato \& Danese (1994).
These and others authors have proposed 
different models for the absorbing material. 
Pier \& Krolik (1992;1993), Granato \& Danese (1994) and Efstathiou \&
Rowan-Robinson (1994) have modeled the absorbing
 structures as axially symmetric tori. 
The models proposed by Granato \& Danese (1994) result in thin and extended 
tori with optical depths
ranging from $\tau \approx$ 10 to 300 in the UV band and maximum 
radii ranging from tens
to hundreds of parsecs. On the contrary, 
the models proposed by Pier \& Krolik (1993)
show thin and compact accretion disks with very large optical
depths with values of $\tau \stackrel{<}{\sim}$ 
1000 in the UV band,
and compact radii with dimension of a few pc. If we consider the Seyfert 1
as canonical unobscured objects, and
abscribe the differences found 
between the two Seyfert types to absorption by the obscuring torus we obtain
a mid IR optical depth of $\tau_{IR} \approx $ 0.4 or a $\tau_{UV} \approx $
80 (assuming $\tau_\lambda \propto \;1/\lambda$) for the Seyfert 2
objects.  This value is indeed very mild and within the range predicted
 for the thin
and extended tori of Granato, Danese \& Franceschini (1997).

The validity of ratioing with the 20~cm radio flux has been however
questioned based on the idea that Seyfert 2 galaxies may have more
star formation in their disks than Seyfert 1s (Maiolino et al. 1995). This
would affect the radio emission, and hence the warm to radio flux ratio. 
It remains to test wether the Seyfert 2
galaxies are indeed stronger emitters of extended far IR radiation. We
consider for this test the ratio of the cold far IR component to 20~cm radio
flux. For the two groups the values are:

\[\begin{array}{lll} 
   \langle  log (F_{cold}/F_{20cm})\rangle_1 &= 6.7 & \sigma=0.7 \\
   \langle  log (F_{cold}/F_{20cm}) \rangle_2 &= 6.5 & \sigma=0.5 \\ 
\end{array}\]\\
 
These distributions are similar (60\% probability) 
according to the KS test. Therefore, Seyfert 2s are
not stronger emitters in extended IR radiation than Seyfert 1s, and there are
no reasons to suspect that they should be stronger radio emitters based
solely on the amount of star formation occurring in Seyfert 2s. It is also worth
pointing out that the differences found between the type 1 and 2 Seyferts are
restricted to the warm emission, while there are not differences regarding
the cold and very cold emission, i.e.,the emission from their respective galaxy disks.

\section{Summary}

We have presented far IR photometry with ISO of the CfA Seyfert sample. The data
have allowed a detailed study of the far IR SED of these
sources using a Bayesian inversion method. It has been
shown that the mid and far IR emission of Seyfert galaxies can be explained
by the emission of three thermal components, a warm component, associated
with dust heated by the nucleus and circumnuclear starformation regions; a
cold dust component heated by star forming region in the galaxy disk, and
very cold dust component heated by the general interstellar radiation
field. The mid to far IR output from Seyfert galaxies does not have a simple
origin but different ingredients play an important role in it.

The comparison of cold far IR fluxes
 with H$\alpha$ data confirms that the cold emission 
component that appeared naturally from the inversion process is 
 related to radiation from dust in starforming regions in the 
galaxy discs.

 We find that the mid IR emission (warm component) is larger in
Seyfert 1 than in 2s, suggesting the presence of obscuring material in
Seyfert 2s. The median value obtained for the optical depth is in the
range predicted by the thin and extended tori models.  


\newpage
\figcaption[FIGURE1.ps]{Spectral Energy Distributions of the CfA Seyfert
sample. Fluxes are in Janskys\label{fig1}}
\figcaption[FIGURE2.ps]{Mid and far IR SEDs of the CfA Seyfert sample. For
each object, the upper panel shows the ISO (triangles) and IRAS (filled
squares) data and the best fit to the SED obtained with the Planckian inverse
transform. The different thermal components contributing to the fit are
printed in dashed lines. The bottom panel shows the temperature spectrum
responsible for these components \label{fig2}}
\figcaption[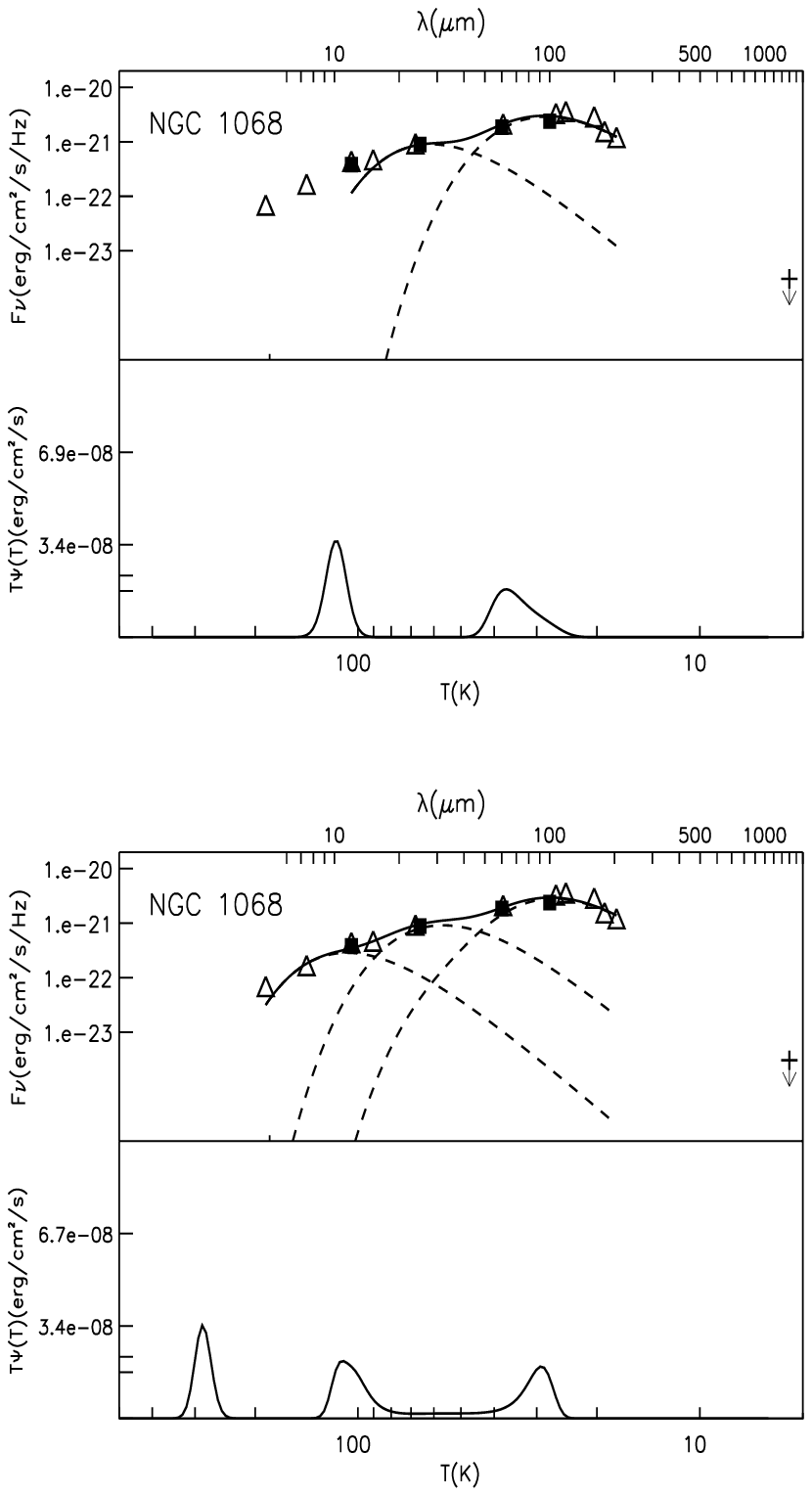]{Mid and far IR 
Spectral Energy Distribution of NGC~1068. As in Figure
2, ISO data are triangles and IRAS data are filled squares. Top: 
The two thermal components obtained with the inversion method are plotted
with dashed lines. In
this case, we have not considered the two first shorwavelength data 
points in order that the
spectral range are equal to the rest of CfA sample. \label{fig3} Bottom:
Result after applying the inversion method to all the complete data set.}
\figcaption[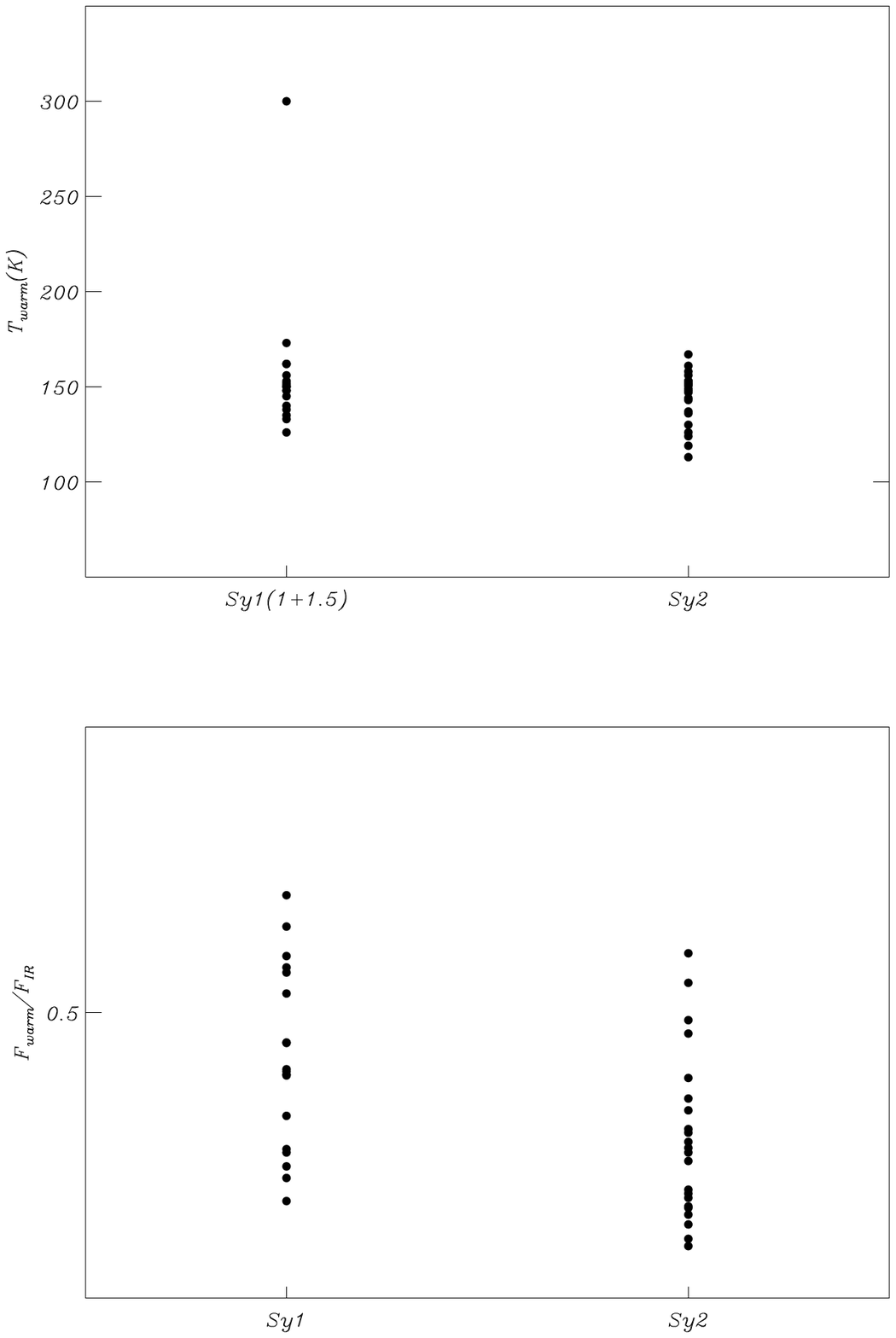]{Top: Distribution of 
temperatures of the warm components for the
objects of the CfA sample. Note that the value of IZw1 is clearly higher than
the rest of the temperatures. \label{fig4}
Bottom: Distribution of the ratio between the warm and the
total fluxes of the CfA sample}
\newpage

\begin{deluxetable}{lcccccccc}
\tablecaption{\label{tab1}ISO fluxes (in Janskys)  for the CfA Seyfert sample}
\tablehead{
\colhead{ Object}&
\colhead{ 16$\mu$m} & 
\colhead{ 25$\mu$m}&
\colhead{ 60$\mu$m}&
\colhead{ 90$\mu$m}&
\colhead{ 120$\mu$m}&
\colhead{ 135$\mu$m}&
\colhead{ 180$\mu$m}&
\colhead{ 200$\mu$m}}
\startdata
Mkr 334 &0.39 & 0.83 & 6.44  & 7.93  & 2.41 & 1.66   & 0.38  & 0.35 \nl
Mkr 335 &0.24 & 0.36 & 0.71  & 0.43  &($^2$) &($^2$) &($^2$)& ($^2$)        \nl
0048+29$^1$ & &                                            	    \nl
IZw 1   &0.64 & 1.08 & 3.42  & 2.43  & 1.30  & 0.79  & 0.45  & ($^2$)      \nl
Mkr 993 &0.13 & 0.09 & 0.34  & 0.77  & 0.47  & 0.17  & 0.43  & 0.39 \nl 
Mkr 573 &0.55 & 0.81 & 3.60  & 1.79  & ($^2$)&($^2$)&  ($^2$) &($^2$)       \nl
0152+06 &0.25 & 0.24 & 1.19  & 1.79  & 4.21  & 4.19  & 1.57  & 1.33 \nl
Mkr 590 &0.46 & 0.26 & 3.06 & 2.59  & ($^3$)&($^3$)&($^3$)&($^3)$        \nl
NGC 1144&0.46 & 0.75 & 8.21  & 14.70 & 22.04  & 22.39&14.22 & 12.04 \nl
Mkr 1243&0.21 & ($^2$) & 0.25  & 0.28  & 0.94  & 1.150& 1.01  & 0.62 \nl
NGC 3079&2.07 & 2.99 & 55.25 & 55.95 & 162.23& 165.96&109.89& 95.12 \nl
NGC 3227&1.22 & 2.07 & 13.57 & 12.35 & 30.45 & 34.38 & 25.87 & 23.43\nl
NGC 3362&0.21 & 0.35 & 2.13  & 3.10  & 3.27  & 3.79  & 2.83  & 3.24 \nl
1058+45 &0.13 & 0.28 & 0.81  & 1.81  & 2.25  & 2.61  & 1.74  & 1.38 \nl
NGC 3516&0.77 & 0.92 & 3.11  & 2.43  & 1.50  & 1.11  & 0.52  & 0.38 \nl
Mkr 744 &0.19 & 0.39 & 1.20  & 1.53  & 5.93  & 6.23  & 4.31  & 3.97 \nl
NGC 3982&0.43 & 0.83 & 8.19  & 11.55 & 19.07 & 20.26 & 17.70 & 14.45\nl
NGC 4051&1.44 & 2.17 & 7.65  & 11.85 & 47.31 & 54.22 & 38.11 & 36.90\nl
NGC 4151&4.12 & 5.18 & 6.28  & 6.94  & 7.90  & 8.52  & 5.73  & 4.69 \nl
NGC 4235&0.17 & 0.28 & 0.65  & 0.66  & 0.19  & 0.74  & 0.57  & 0.30 \nl
Mkr 766 &0.73 & 1.35 & 6.05  & 5.13  &3.74   & 3.47  & 2.72  & 2.10 \nl
Mkr 205 &0.19 & 0.27 & 0.69  & 1.25  & 2.24  & 3.00  & 2.64  & 3.20 \nl
NGC 4388&1.55 & 3.40 & 11.63 & 16.44 & 24.28 & 28.20 & 19.07 & 16.78\nl
Mkr 231 &3.96 & 6.90 & 76.27 & 61.45 & 37.72 & 26.67 & 15.97 & 13.34\nl
NGC 5033&1.34 & 1.78 & 18.77 & 32.96 & 93.13 & 104.08& 75.87 & 66.93\nl
Mrk 789$^1$   &       &                                             \nl 
1335+39 &0.13 & 0.24 & 1.80  & 2.39  & 2.39  & 2.95  & 1.78  & 1.64 \nl
NGC 5252&0.10 & 0.14 & 1.85  & 1.21  & 0.97  & 1.50  & 0.70  & 0.53 \nl
Mkr 266 &0.33 & 1.07 & 9.72  & 11.11 & 11.60 & 9.41  & 6.98  & 5.11 \nl
Mkr 270 &0.06 & 0.13 & 0.21  & 0.09  & 0.64  & 0.57  & 0.43  & 0.62 \nl
NGC 5273&0.09 & 0.18 & 1.24  & 1.95  & 1.18  & 1.19  & 0.96  & 0.63 \nl
Mkr 461 &($^2$) & 0 20 & 0.15  & 0.32  & 0.25  & 0.30  & 0.21  & 0.30 \nl
Mkr 279 &0.29 & 0.50 & 1.58  & 2.20  & 2.54  & 2.62  & 1.46  & 1.59 \nl
IC 4397 &0.16 & 0.22 & 2.43  & 3.69  & 4.17  & 4.36  & 2.82  & 1.65 \nl
NGC 5548&0.44 & 0.79 & 1.23  & 1.25  & 1.14  & 0.95  & 0.57  & 0.72 \nl
NGC 5674$^1$  &                                    		    \nl
Mkr 817 &0.49 & 1.11 & 4.71  & 3.80  & 1.49  & 1.37  & 0.62  & 0.69 \nl
Mkr 686$^1$ &     &                                                     \nl
Mkr 841 &0.34 & 0.50 & 0.57  & 0.29  &($^2$) &($^2$)&($^2$)&($^2$)          \nl
NGC 5929&0.55 & 1.50 & 13.11 & 16.01 & 19.50 & 17.32 & 10.33 & 7.39 \nl
NGC 5940$^1$  & &                                                   \nl 
1614+35 &0.12 & 0.16 & 0.76  &  0.48 & 1.73  & 1.79  & 1.41  & 1.50 \nl
2237+07 &0.25 & 0.35 & 0.50  &  0.54 & 0.70  & 0.71  & 0.51  & 0.22 \nl
NGC 7469&1.77 & 4.78 & 33.56 & 38.63 & 46.23 & 42.26 & 28.92 & 22.32\nl
Mkr 530$^1$   &       &                                             \nl 
Mkr 533 &0.86 & 1.75 & 4.58  & 6.32  & 7.63  & 7.61  & 6.63  & 3.75 \nl
NGC 7682&0.29 & 0.22 & 0.47  & 0.30  & 0.63  & 0.78  & 0.54  & 0.66 \nl
\tablecomments{$^1$ Not observed\nl
$^2$ Not detected\nl
$^3$ Background not available}
\enddata
\end{deluxetable}
\newpage

\begin{deluxetable}{llllllllllll}
\tablecaption{Fluxes of NGC 1068. Fluxes are
given in Janskys \label{tab2}}
\tablehead{
$\lambda (\mu m)$&
\colhead{ 4.9}&
\colhead{ 7.6}&
\colhead{ 11.5}&
\colhead{ 16}& 
\colhead{ 25}&
\colhead{ 60}&
\colhead{ 105}&
\colhead{ 120}&
\colhead{ 135}&
\colhead{ 180}&
\colhead{ 200}} 
\startdata
{ F (Jy)}&6.9&16.7&44.0&46.9&92.6&212.4&331.2&360.6&288.3&155.2&120.8\nl
\enddata
\end{deluxetable}

\newpage

\begin{deluxetable}{cccccc}
\tablecaption{\label{tab3}Central temperature of spectral components}
\tablehead{
\colhead{Object}&
\colhead{ Type }&     
\colhead{Morph. type}&
\colhead{T$_1$(K)} &  
\colhead{T$_2$(K)} &  
\colhead{T$_3$(K)}
}
\startdata
Mrk 334  &  2 & Pec.&    & 46 & 151  \nl
Mrk 335  &  1 & S0/a& 32 & 71 & 162  \nl
IZw1     &  1 & S?  & 40 & 97 & $\sim$300 \nl
Mrk 993  & 1.5& Sa  & 30 & 65 & 173  \nl
Mrk 573  &  2 & SAB0& 35 & 75 & 143  \nl
0152+06  &  2 & SAb & 22 & 43 & 153  \nl
Mrk 590$^2$ &  1 & SAa &  &  &   \nl
NGC 1068$^1$&  2 & SAb & &36 & 115 \nl
NGC 1144 &  2 & R   & 22 & 50 & 148   \nl
Mrk 1243 &  1 & Sa  & 19 & 51 & 138   \nl
NGC 3079 &  L & SBc & 22 & 42 & 152   \nl
NGC 3227 & 1.5& SABa& 19 & 43 & 126   \nl
NGC 3362 &  2 & SABc& 20 & 42 & 113   \nl
1058+45  &  2 & Sa  & 22 & 53 & 144   \nl
NGC 3516 & 1.5& SB0 &    & 42 & 148   \nl
Mrk 744  &  2 & SABa& 18 & 55 & 137   \nl
NGC 3982 &  2 & SABb& 20 & 42 & 124   \nl
NGC 4051 &  1 & SABbc & 20 & 49 & 148 \nl
NGC 4151 &  1 & SABab & 21 & 53 & 151 \nl
NGC 4235 &  1 & SAa & 16 & 38 & 145   \nl
Mrk 766  & 1.5& SBa & 33 & 56 & 133   \nl
Mrk 205  &  1 & SBab& 16 & 28 & 156   \nl
NGC 4388 &  2 & SAb & 21 & 43 & 119   \nl
Mrk 231  &  1 & SAc &    & 39 & 150   \nl
NGC 5033 &  2 & SAc & 21 & 44 & 147   \nl
1335+39  &  2 & S?  & 28 & 53 & 156   \nl
NGC 5252 &  2 & S0  & 20 & 56 & 167   \nl
Mrk 266  &  2 & Pec.& 27 & 45 & 130   \nl
Mrk 270  &  2 & S0  & 18 & 55 & 161   \nl
NGC 5273 &  1 & SA0 & 31 & 63 & 151   \nl
Mrk 461  &  2 & S   & 20 & 63 & 153   \nl
Mrk 279  &  1 & S0  & 24 & 46 & 153   \nl
IC 4397  &  2 & S?  & 26 & 46 & 158   \nl
NGC 5548 &  1 & SA0 & 23 & 51 & 152   \nl
Mrk 817  &  1 & S?  &    & 45 & 135   \nl
Mrk 841  & 1.5& E   &    & 71 & 162   \nl
NGC 5929 &  2 & Sab & 28 & 54 & 126   \nl
1614+35  & 1.5& S?  & 19 & 41 & 150   \nl
2237+07  &  2 & SBa & 22 & 58 & 151   \nl
NGC 7469 &  1 & SABa& 27 & 60 & 140   \nl
Mrk 533  &  2 & SAbc& 21 & 54 & 149   \nl
NGC 7682 &  2 & SBab& 16 & 45 & 136   \nl
\enddata
\tablecomments{
$^1$ Short wavelength data (5-11~$\mu$m) not taken into account (See section
\ref{n1068})}
\end{deluxetable}

\begin{deluxetable}{ccllll}
\tablecaption{\label{tab4}Fluxes for each component of the SEDs (erg
cm$^{-2}$s$^{-1}$). F$_1$ corresponds to the {\bf very cold} component, F$_2$
to the {\bf cold} component and F$_3$ to the {\bf warm
component}. F$_{total}$ is the total sum of the fluxes}
\tablehead{
\colhead{Object}&
\colhead{ Type }&     
\colhead{F$_1$} &  
\colhead{F$_2$} &  
\colhead{F$_3$}&
\colhead{F$total$}
}
\startdata
Mrk 334 & 2 &  &2.964 10$^{-10}$ &9.618  10$^{-11}$   & 3.655 10$^{-10}$\nl
Mrk 335 & 1 & 1.326 10$^{-11}$ &2.357 10$^{-11}$ &4.211 10$^{-11}$    &7.894
10$^{-11}$  \nl
IZw1    & 1 & 1.135 10$^{-10}$ &9.073 10$^{-11}$ &$>$1.345 10$^{-10}$ & $>$3.453 10$^{-10}$ \nl
Mrk 993 &1.5& 2.072 10$^{-11}$ &0.793 10$^{-11}$ &2.322 10$^{-11}$    & 5.187 10$^{-11}$\nl
Mrk 573 & 2 & 9.210 10$^{-11}$ &2.596 10$^{-11}$ &1.120 10$^{-10}$    & 2.300 10$^{-10}$\nl
0152+06 & 2 & 7.836 10$^{-11}$ &1.823 10$^{-11}$ &4.062 10$^{-11}$    & 1.372
10$^{-10}$\nl
Mrk 590$^2$  &  1 &&  &  &   \nl
NGC 1068& 2 & & 1.170 10$^{-8}$  &1.198 10$^{-8}$  \nl
NGC 1144& 2 &5.131 10$^{-10}$ &1.702 10$^{-10}$ &1.012 10$ ^{-10}$	 & 7.845 10$^{-10}$\nl
Mrk 1243& 1 &2.187 10$^{-11}$ &1.209 10$^{-11}$ &2.188 10$ ^{-11}$$^*$ & 5.584 10$^{-11}$\nl
NGC 3079& L &2.929 10$^{-9}$ &1.344 10$^{-9}$  &4.283 10$ ^{-10}$	 & 4.701 10$^{-9}$\nl
NGC 3227&1.5&5.505 10$^{-10}$ &3.850 10$^{-10}$ &2.805 10$ ^{-10}$	 & 1.216 10$^{-9}$\nl
NGC 3362& 2 &5.168 10$^{-11}$ &9.518 10$^{-11}$ &5.533 10$ ^{-11}$$^*$ & 2.022 10$^{-10}$\nl
1058+45 & 2 &5.712 10$^{-11}$ &2.351 10$^{-11}$ &3.289 10$ ^{-11}$	 & 1.135 10$^{-10}$\nl
NGC 3516&1.5&	   &1.201 10$ ^{-10}$&1.593 10$ ^{-10}$	  & 2.794 10$^{-10}$\nl
Mrk 744 & 2 &1.923 10$^{-10}$ &5.239 10$^{-11}$ &4.684 10$ ^{-11}$	 & 2.924 10$^{-10}$\nl
NGC 3982& 2 &3.597 10$^{-10}$ &2.951 10$^{-10}$ &1.121 10$ ^{-10}$	 & 7.669 10$^{-10}$\nl
NGC 4051& 1 &8.771 10$^{-10}$&2.690 10$^{-10}$ &3.053 10$ ^{-10}$	 & 1.1.451 10$^{-9}$\nl
NGC 4151& 1 &1.519 10$^{-10}$&2.961 10$^{-10}$ &8.346 10$ ^{-10}$	 & 1.282 10$^{-9}$\nl
NGC 4235& 1 &3.132 10$^{-12}$ &2.557 10$^{-11}$ &3.950 10$ ^{-11}$	 & 6.820 10$^{-11}$\nl
Mrk 766 &1.5&1.537 10$^{-10}$ &1.187 10$^{-10}$ &1.741 10$ ^{-10}$	 & 4.465 10$^{-10}$\nl
Mrk 205 & 1 &3.669 10$^{-11}$ &3.464 10$^{-11}$ &3.344 10$ ^{-11}$	& 1.047 10$^{-10}$\nl
NGC 4388& 2 &4.980 10$^{-10}$ &4.286 10$^{-10}$ &4.540 10$ ^{-10}$	 & 1.380 10$^{-9}$\nl
Mrk 231 & 1 &	   &2.711 10$ ^{-9}$ &9.569 10$ ^{-10}$	  & 3.668 10$^{-9}$\nl
NGC 5033& 2 &1.822 10$^{-9}$  &4.121 10$^{-10}$ &2.586 10$ ^{-10}$	 &
2.502 10$^{-9}$\nl
1335+39 & 2 &8.249 10$^{-11}$ &2.323 10$^{-11}$ &3.340 10$ ^{-11}$	 & 1.391 10$^{-10}$\nl
NGC 5252& 2 &2.250 10$^{-11}$ &2.232 10$^{-11}$ &2.812 10$ ^{-11}$	 & 7.294 10$^{-11}$\nl
Mrk 266 & 2 &2.049 10$^{-10}$ &3.164 10$^{-10}$ &1.168 10$ ^{-10}$	 & 6.381 10$^{-10}$\nl
Mrk 270 & 2 &1.244 10$^{-11}$ &5.686 10$^{-12}$ &1.565 10$ ^{-11}$     & 3.378 10$^{-11}$\nl
NGC 5273& 1 &5.174 10$^{-11}$ &3.219 10$^{-11}$ &1.576 10$ ^{-11}$	 & 9.969 10$^{-11}$\nl
Mrk 461 & 2 &5.384 10$^{-12}$ &1.874 10$^{-11}$ &2.974 10$ ^{-11}$	& 5.386 10$^{-11}$\nl
Mrk 279 & 1 &7.046 10$^{-11}$ &3.165 10$^{-11}$ &6.831 10$ ^{-11}$	 & 1.704 10$^{-10}$\nl
IC 4397 & 2 &1.194 10$^{-10}$ &3.566 10$^{-11}$ &3.635 10$ ^{-11}$	 & 1.913 10$^{-10}$\nl
NGC 5548& 1 &1.769 10$^{-11}$ &5.768 10$^{-11}$ &1.126 10$ ^{-10}$	 &
1.898 10$^{-10}$\nl
Mrk 817 & 1 &	   &1.680 10$ ^{-10}$&1.357 10$ ^{-10}$	  & 3.037 10$^{-10}$\nl
Mrk 841 &1.5&	   &2.913 10$ ^{-11}$&6,982 10$ ^{-11}$	 & 9.895 10$^{-11}$\nl
NGC 5929& 2 &5.487 10$^{-10}$ &2.282 10$^{-10}$ &1.654 10$ ^{-10}$	 & 9.423 10$^{-10}$\nl
1614+35 &1.5&3.392 10$^{-11}$ &2.227 10$^{-11}$ &1.923 10$ ^{-11}$	 & 7.542 10$^{-11}$\nl
2237+07 & 2 &1.297 10$^{-11}$ &3.840 10$^{-11}$ &5.347 10$ ^{-11}$	 & 1.048 10$^{-10}$\nl
NGC 7469& 1 &1.231 10$^{-9}$  &9.128 10$^{-10}$ &4.390 10$ ^{-10}$	 & 2.583 10$^{-9}$\nl
Mrk 533 & 2 &1.532 10$^{-10}$ &2.594 10$^{-10}$ &2.217 10$ ^{-10}$	 & 6.643 10$^{-10}$\nl
NGC 7682& 2 &1.058 10$^{-11}$ &1.796 10$^{-11}$ &4.352 10$ ^{-11}$	 & 7.206 10$^{-11}$\nl
\enddata
\end{deluxetable}
\begin{deluxetable}{ccllll}
\tablecaption{\label{tab5}Luminosities for each component of the SEDs
(10$^{44}$erg
s$^{-1}$). L$_1$ corresponds to the {\bf very cold} component, L$_2$
to the {\bf cold} component and L$_3$ to the {\bf warm
component}. L$_{IR}$ is the total luminosity}
\tablehead{
\colhead{Object}&
\colhead{ Type }&     
\colhead{L$_1$} &  
\colhead{L$_2$} &  
\colhead{L$_3$}&
\colhead{L$_{IR}$}
}
\startdata
Mrk 334  &  2 &      & 25.19 & 9.00 & 34.19\nl
Mrk 335  &  1 & 1.72 & 3.06  & 5.47 & 10.25\nl
IZw1     &  1 & 81.49& 65.14 & $>$96.57 &$>$243.20\nl
Mrk 993  & 1.5& 0.94 & 0.36  & 1.06 & 3.76\nl
Mrk 573  &  2 & 5.31 & 1.50  & 6.46 & 13.27\nl
0152+06  &  2 & 4.57 & 1.06  & 2.37 & 8.10\nl
Mrk 590$^2$&1 &\nl
NGC 1068 &  2 &      & 30.06 &31.93 & 61.99\nl
NGC 1144 &  2 & 82.51& 27.37 &16.27 & 126.15\nl
Mrk 1243 &  1 & 5.30 & 2.93  &5.30  & 13.53\nl
NGC 3079 &  L & 7.68 & 3.52  &1.12  & 12.32\nl
NGC 3227 & 1.5& 1.52 & 1.06  &0.77  & 3.35\nl
NGC 3362 &  2 & 5.15 & 9.48  &5.51  & 20.14\nl
1058+45  &  2 & 9.38 & 3.86  &5.40  & 18.64\nl
NGC 3516 & 1.5&      & 1.67  &2.21  & 3.88\nl
Mrk 744  &  2 & 3.06 & 0.83  &0.75  & 4.64\nl
NGC 3982 &  2 & 0.62 & 0.51  &0.19  & 1.32\nl
NGC 4051 &  1 & 0.81 & 0.25  &0.28  & 1.34\nl
NGC 4151 &  1 & 0.26 & 0.51  &1.44  & 2.21\nl
NGC 4235 &  1 & 0.035& 0.29  &0.45  & 0.77\nl
Mrk 766  & 1.5& 4.84 & 3.74  &5.49  & 14.07\nl
Mrk 205  &  1 &35.53 &33.55  &32.39 & 101.47\nl
NGC 4388 &  2 &6.12  &5.26   &5.58  & 16.96\nl
Mrk 231  &  1 &      &888.77  &313.71& 1202.48\nl
NGC 5033 &  2 &3.14  &0.71   &0.44  & 4.29\nl
1335+39  &  2 &6.43  &18.12  &26.05 & 50.60\nl
NGC 5252 &  2 &23.20  &23.02  &29.01& 75.23\nl
Mrk 266  &  2 &30.02 &46.37  &17.11 & 93.50\nl
Mrk 270  &  2 &0.19  &0.09  &0.24   & 0.52 \nl
NGC 5273 &  1 &0.13  &0.08  &0.04   & 0.25 \nl
Mrk 461  &  2 &0.27  &0.92   &1.47  & 2.66\nl
Mrk 279  &  1 &12.63 &5.68   &12.25 & 30.56\nl
IC 4397  &  2 &4.97  &1.48   &1.51  & 7.96 \nl
NGC 5548 &  1 &0.94  &3.06   &5.98  & 9.98 \nl
Mrk 817  &  1 &      &32.15  &25.97 & 58.12\nl
Mrk 841  & 1.5&      &7.51   &18.00 & 25.51\nl
NGC 5929 &  2 &7.25  &3.02   &2.19  & 12.46\nl
1614+35  & 1.5&5.15  &3.38   &2.92  & 11.45\nl
2237+07  &  2 &1.57  &4.64   &6.47  & 12.68\nl
NGC 7469 &  1 &60.72 &45.02  &21.65 & 127.39\nl
Mrk 533  &  2 &24.81 &42.01  &35.90 & 102.72\nl
NGC 7682 &  2 &0.59  &1.00   &2.42  & 4.01\nl
\enddata

\end{deluxetable}


\begin{thebibliography}{}

 \bibitem{}  Antonucci, R.J.J., Miller, J.S., 1985, ApJ, 297, 621
 \bibitem{} Armus, L., Heckman, T.M., Miley, G.K., 1990, \apj, 364, 471
 \bibitem{}  Bonatto, C.J., Pastoriza, M.G., 1997, ApJ, 486, 132
 \bibitem{}  Boroson, T.A., Meyers, K.A., 1992, ApJS, 80, 109
 \bibitem{}  Braatz, J.A., Wilson, A.S., Gezari, E.Y., Varosi, J., Beichman,
 C.A., 1993, ApJ, 409, L5
 \bibitem{}  Bregman, J.N., 1990, A\&A Rev, 2, 125
 \bibitem{}  Cameron, M., Storey, J.W.V., Rotaciuc, V., Genzel, R.,
 Verstraete, L., Drapatz, S., Siebenmorgen, R., Lee, T.J., 1993, ApJ, 419,
 136
 \bibitem{}  Chini, R., Kruger, E.,  Kreysa, E., 1992, A\&A, 315, 75
 \bibitem{}  Cox, R., Kruger, E., Mezger, P.G., 1986, A\&A,, 155, 380 
 \bibitem{}  Danese, L., Zitelli, V., Granato, G.L., Wade, R., De Zotti, G.,
 Mandolesi, N., 1992, ApJ, 399, 38
 \bibitem{} Devereux, N.A., Young, J.S., 1990, \apj, 350, L25
 \bibitem{}  Edelson, R.A., 1987, ApJ, 313, 651
 \bibitem{}  Edelson, R.A., Malkan, M.A., Rieke, G.H., 1987, ApJ, 321, 233
 \bibitem{}  Efstathiou, A., Rowan-Robinson, M., 1994, \mnras, 212, 218
 \bibitem{}  Giuricin, G., Mardirossian, F., Mezzetti, M., 1995, ApJ, 446,
 550
 \bibitem{} Gonz\'alez Delgado, R.M., P\'erez, E., Tadhunter, C., 
Vilchez, J.M., Rodr\'{\i}guez Espinosa, J.M.,  1997, \apjs, 108, 155
 \bibitem{}  Gonz\'alez Delgado, R.M., Heckman, T., Leitherer, C., Meurer,
 G., Krolik, L., Wilson, A., Kinney, A., Koratkar, A., 1998, ApJ, 505, 174
 \bibitem{}  Goodrich, R.W., 1989, ApJ, 342, 234
 \bibitem{}  Granato, G.L., Danese, L., 1994, MNRAS, 268, 233
 \bibitem{}  Granato, G.L., Danese, L., Franceschini, A., 1997, \apj, 487, 147
 \bibitem{}  Halpern, J.P., Moran, E.C., 1998, ApJ, 494, 194
 \bibitem{} Hawarden, T.G., Israel, F.P., Geballe, T.R., Wade, R., 1995,
 \mnras, 276, 119
 \bibitem{}  Heckman, T.M., 1995, ApJ, 446, 101
 \bibitem{}  Heckman, T.M., Gonz\'alez-Delgado, R.M., Leitherer, C., Meurer,
 G.R., Krolik, J., Wilson, A.S., Koratkar, A., Kinney, A., 1997, ApJ, 482,
 114
 \bibitem{}  Huchra, J., Burg, R., 1992, ApJ, 393, 90
 \bibitem{} Kennicutt, R.C. Jr., 1983, \apj, 272, 54
 \bibitem{} Kessler, M. F., Steinz, J.A., Anderegg, M.E., Clavel, J.,
 Drechsel, G., Estaria, P., Faelker, J., Riedinger, J.R., Robson, A., Taylor,
 B.G., Ximenez de Ferran, S., 1996, \aap, 315, L27
 \bibitem{}  Klaas, U., Haas, M., Heinrichsenm, J., Schulz, B., 1997, A\&A,
 325, L21
 \bibitem{}  Knapp, G.R., Rupen, M.P., Fich, M., Harper, D.A., Wynn-Williams,
 C.G., 1996, A\&A, 315, L75
 \bibitem{}  Lawrence, A., Elvis, M., Wilkes, B.J., McHardy, I., Brandt, N.,
 1997, MNRAS, 285, 879
 \bibitem{}  Lemke, D., et al. 1996, A\&A, 315, L64
 \bibitem{}  Lonsdale, C.J., Helou, G., Good, J.C., Rice, W.,1984,
 ``Cataloged galaxies and quasars observed in the IRAS survey''
 \bibitem{}  Maiolino, R., Krabbe, A., Thatte, N., Genzel, R., 1998, ApJ,
 493, 650
 \bibitem{}  Maiolino, R., Ruiz, M., Rieke, G.H., Keller, L.D., 1995, ApJ,
 446, 101
 \bibitem{}  Miller, J.S., Antonucci, R.J.J., 1983, ApJ, 271, 7
 \bibitem{}  Mulchaey, J.S., Koratkar, A., Ward, M.J., Wilson, A.S., Whittle,
 M., Antonucci, R.J.J., Kinney, A.L., Hurt, T., 1994, ApJ, 436, 658
 \bibitem{}  Neugebauer, G., 1978, PhyS, 17, 149
 \bibitem{}  Osterbrock, D.E., Pogge, R.W., 1985, ApJ, 297, 166
 \bibitem{}  Prieto, M.A., P\'erez Garc\'{\i}a, A.M., Rodr\'{\i}guez
 Espinosa, J.M., 1999 (in preparation)
 \bibitem{}  P\'erez Garc\'{\i}a, A.M., Rodr\'{\i}guez Espinosa, J.M.,
 Fuensalida ,J.J., 1998, ApJ (submitted) 
 \bibitem{}  P\'erez Garc\'{\i}a, A.M., Rodr\'{\i}guez Espinosa,
 J.M., Santolaya Rey A.E., 1998, ApJ, 500, 685
 \bibitem{}  Pier, E.A., Krolik, J.H., 1993, ApJ, 418, 673
 \bibitem{}  Pier, E.A., Krolik, J.H., 1992, ApJ, 401, 99
 \bibitem{} Rieke, G.H., 1978, ApJ, 226, 550
 \bibitem{} Rieke, G.H., Low, F.J., 1972, ApJ, 176, L95 
 \bibitem{} Rigopoulou, D., Papadakis, I., Lawrence, A., Ward, M., 1997,
 A\&A, 327, 493
 \bibitem{} Rodr\'{\i}guez Espinosa, J.M., P\'erez Garc\'{\i}a, A.M.,
 Lemke, D., Meisenheimer, K., 1996, A\&A, 315, L129
 \bibitem{} Rodr\'{\i}guez Espinosa, J.M., P\'erez Garc\'{\i}a, A.M., 1997,
 ApJ, 487, L33
 \bibitem{} Rodr\'{\i}guez Espinosa, J.M., Rudy, R.J., Jones, B., 1987, ApJ,
 312, 555
 \bibitem{} Romanishin, W., 1990, \aj, 100, 373
 \bibitem{} Rudy, R.J., 1984, ApJ, 284, 33
 \bibitem{} Salas, L., 1992, ApJ, 385, 288
 \bibitem{} Siebenmorgen, R., Moorwood, A., Friedling, W., Kaeufl, H.U.,
 1997, A\&A, 325, 450
 \bibitem{} Simpson, C., Wilson, A.S., Bower, G., Heckman, T.M., Krolik,
 J.H., Miley, G.K., 1997, ApJ, 474, 121
 \bibitem{} Smith, M.A., Lada, C.J., Thronson, H.A., Glaccum, W., Harper,
 D.W., Loewenstein, R.F., Smith, J., 1983, ApJ, 274, 571
 \bibitem{} Spinoglio, I., Malkan, M.A., Rush, B., Carrasco, L.,
 Recillas-Cruz, E., 1995, ApJ, 453, 616
 \bibitem{} Stein, W.A., 1975, PASP, 87, 5 
 \bibitem{} Telesco, C.M., 1990, in ``Infrared Astronomy'', Mampaso,
 A. Prieto, M., S\'anchez, F., Eds., Cambridge University Press, p173
 \bibitem{} Telesco, C.M., 1988, ARA\&A, 26, 343
 \bibitem{} Telesco, C.M., Harper, D.A., 1980, ApJ, 235, 392 
 \bibitem{} Tresh-Frenberg, R., Fazio, G.G., Gezari, D.Y., Lamb, G.M., Shu,
 P.K., Hoffmann, W.F., McReight, C.R., 1987, ApJ, 312, 542
 \bibitem{} Walterbos, R.A.M., Greenwalt, B., 1996, ApJ, 460, 696 
 \bibitem{} Walterbos, R.A.M., Schwering, P.B.W., 1987, A\&A, 180, 27
 \bibitem{} Wilson, A.S., Braatz, J.A., Heckman, T.M., Krolik, J.H., Miley,
 G.K., 1993, ApJ, 419, L61
 \bibitem{} Young, J.S., Xie, S., Kenney, J.D.P., Rice, W.L., 1989, ApJS, 70,
 699
\end{thebibliography}
\end{document}